\documentclass[10pt,conference]{IEEEtran}

\usepackage{times}
\usepackage{graphicx}
\usepackage{multirow}
\usepackage{subfigure}
\usepackage{url}
\usepackage{amssymb}
\usepackage{dsfont}

\begin{document}

\title{Fundamental Delay Bounds in Peer-to-Peer Chunk-based Real-time Streaming Systems}

\author{\IEEEauthorblockN{
            Giuseppe Bianchi,
            Nicola Blefari Melazzi,
            Lorenzo Bracciale,
            Francesca Lo Piccolo,
            Stefano Salsano
            }
\IEEEauthorblockA{
            Dipartimento di Ingegneria Elettronica,
            Universit\`a di Roma Tor Vergata\\
            \{giuseppe.bianchi,blefari,lorenzo.bracciale,francesca.lopiccolo,stefano.salsano\}@uniroma2.it}
            }

\maketitle

\begin{abstract}
This paper addresses the following foundational question: what is
the maximum theoretical delay performance achievable by an overlay
peer-to-peer streaming system where the streamed content is
subdivided into chunks? As shown in this paper, when posed for
chunk-based systems, and as a consequence of the store-and-forward
way in which chunks are delivered across the network, this question
has a fundamentally different answer with respect to the case of
systems where the streamed content is distributed through one or
more flows (sub-streams). To circumvent the complexity emerging when
directly dealing with delay, we express performance in term of a
convenient metric, called ``stream diffusion metric''. We show that
it is directly related to the end-to-end minimum delay achievable in
a P2P streaming network. In a homogeneous scenario, we derive a
performance bound for such metric, and we show how this bound
relates to two fundamental parameters: the upload bandwidth
available at each node, and the number of neighbors a node may
deliver chunks to. In this bound, k-step Fibonacci sequences do
emerge, and appear to set the fundamental laws that characterize the
optimal operation of chunk-based systems. 
\end{abstract}

\section{Introduction}
\label{s:intro}

Nowadays, Peer-to-Peer (P2P) overlay live streaming systems are of
significant interest, thanks to their low implementation complexity,
scalability and reliability properties, and ease of deployment.
Leveraging on the well understood P2P communication paradigm, the
viability to deliver live streaming content on top of a
self-organizing P2P architecture has been widely assessed both in
terms of research contributions, as well as in terms of real-life
applications.

In principle, the most natural and earlier solution for deploying a
P2P streaming system was to organize peer nodes in one or more
overlay multicast trees, and hence continuously deliver the streamed
information across the formed paths. This is the case in
\cite{nice,zigzag,splitstream}. 
However, in practice, this approach may not be viable in large-scale
systems and with nodes characterized by intermittent connectivity
(churn). In fact, whenever a node in the middle of a path abruptly
disconnects, complex procedures would be necessary to i) allow the
reconstruction of the distribution path, and ii) allow the nodes
affected by such event to recover the amount of information lost
during the path reconfiguration phases. To overcome such
limitations, a completely different approach, called {\em
data-driven}, delivers content on the basis of content availability
information, locally exchanged among connected peers, without any a
priori pre-established path. This approach essentially creates a
mesh topology among overlay nodes. Several proposed solutions, such
as \cite{coolstreaming,gridmedia-II,prime,pulse}, adopt the
data-driven approach.

In this paper we focus on {\em chunk-based} systems, where,
similarly to most file-sharing P2P applications, the streaming
content is segmented into smaller pieces of information called
chunks. Chunks are elementary data units handled by the nodes
composing the network in a store-and-forward fashion. A relaying
node can start distributing a chunk only when it has completed its
reception from another node. While the solutions based on multicast
overlay trees usually organize the information in form of small IP
packets to be sequentially delivered across the trees and can not be
regarded as chunk-based, some data-driven solutions, like the ones
proposed in \cite{coolstreaming,prime,pulse}, may be regarded as
chunk-based. A characterizing feature of the chunk-based approach is
that, in order to reduce the per-chunk signalling burden, the chunk
size is typically kept to a fairly large value, greater than the
typical packet size.

In this paper we raise some very basic and foundational questions on
chunk-based systems: what are the theoretical performance limits,
with specific attention to delay, that {\em any} chunk-based
peer-to-peer streaming system is bounded to? Which fundamental laws
describe how performances depend on network parameters such as the
available bandwidth or system parameters such as the number of nodes
a peer may at most connect to? And which are the system topologies
and operations which would allow to approach such bounds?

The aim of this paper is to answer these questions. The answer is
completely different from the case of systems where the streaming
information, optionally organized in sub-streams, is continuously
delivered across overlay paths (for a theoretical investigation of
such class of approaches refer to \cite{sigmetrics08} and references
therein contained). As we will show, in our scenario the time needed
for a chunk to be forwarded across a node significantly affects
delay performance.

In more detail, we focus on the ability to reach the greatest
possible number of nodes in a given time interval (this will be
later on formally defined as ``stream diffusion metric'') or
equivalently the ability to reach a given number of nodes in the
smallest possible time interval (i.e. absolute delay). We derive
analytic expressions for the maximum asymptotic stream diffusion
metric in an homogeneous network composed of stable nodes whose
upload bandwidth is the same (for simplicity, multiple of the
streaming rate).

With reference to such homogeneous and ideal scenario, we show how
this bound relates to two fundamental parameters: the upload
bandwidth available at each node, and the number of neighbors a node
may deliver chunks to. In addition, we show that the serialization
of chunk transmissions and the organization of peer nodes into
multiple overlay unbalanced trees allow to achieve the proposed
bound. This suggests that the design of real-world applications
could be driven by two simple basic principles: i) the serialization
of chunk transmissions, and ii) the organization of chunks in
different groups so that chunks in different groups are spread
according to different paths. As a matter of fact, in a companion
paper \cite{O-streamline}, we have indeed presented a simple
data-driven heuristic, called {\em O-Streamline}, which exploits the
idea of using serial transmissions over multiple paths and relies on
a pure data-oriented operation (i.e. chunk paths are not
pre-established). Such heuristic successfully achieves performances
close to the ones of the theoretical bound.

This paper is organized as follows. Section \ref{s:moti} explains
the rational behind this work. Section \ref{s:bound} introduces the
stream diffusion metric and derives the relative bound. The overlay
topology that allows to achieve the presented bound is described in
section \ref{s:algo}. Sections \ref{s:perfo} presents some
performance evaluation results. Section \ref{s:related} reviews the
related work. Finally, section \ref{s:conclu} concludes the paper.

\section{Motivation}
\label{s:moti}%
Goal of this section is to clarify why P2P {\em chunk-based}
streaming systems have significantly different performance issues
with respect to streaming systems, where the information content
continuously flows across one or more overlay paths or trees. Unless
ambiguity occurs, such systems will be referred to as, with slight
abuse of name, {\em flow-based} systems. More precisely, we will
show that i) theoretical bounds derived for the flow-based case may
not be representative for chunk-based systems, and new, {\em
fundamentally different}, bounds are needed, ii) the methodological
approaches which are applicable in the two cases are completely
diverse, and fluidic approaches may be replaced with inherently
discrete-time approaches where, as shown later on, $k$-step
Fibonacci series and sums enter
into play. 

\subsection{Delay in flow-based systems}
\label{ss:flow}

We recall that ``flow-based'' system denotes a stream distribution
approach where the streaming information, possibly organized in
multiple sub-streams, is delivered with continuity across one or
more overlay network paths. Clearly, in the real IP world,
continuous delivery is an abstraction, as the streaming information
will be delivered in the form of IP packets. However, the small size
of IP packets yields marginal transmission times at each node. As
such, the remaining components that cause delay over an overlay link
(propagation and path delay because of queueing in the underlying
network path) may be considered predominant. We can conclude that
the delay performances of flow-based systems ultimately depend on
the delay characterizing a path between the source node and a
generic end-peer. More specifically, if we associate a delay figure
to each overlay link, then the source to destination delay depends
on the sum of the link delays: the transmission times needed by the
flow to ``cross'' a node may be neglected, or, more precisely, they
play a role only because the `crossed'' nodes compose the vertices
of the overlay links, whose delays dominate the overall delay
performance.

\begin{figure}[t]
    \centering
    \includegraphics[scale=0.65]{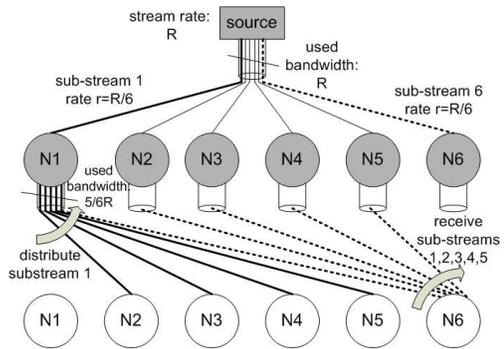}
    \caption{Tree depth optimization in flow-based systems. A tree depth
    equal to 2 can be achieved by i) splitting the stream in a number of
    sub-streams equal to the number of network nodes $N$, ii) delivering each sub-stream
    to a different node, and iii) letting each node $i$
    replicate and deliver the $i$-th sub-stream to the remaining $N-1$ nodes.
    \vspace{-0.35cm}}
    \label{fig:2a}
\end{figure}

As a consequence, the delay performance optimization becomes a
minimum path cost problem, as such addressed with relevant
analytical techniques. If we further assume that the network links
are homogeneous (i.e. characterized by the same delay), then the
problem of finding a delay performance bound is equivalent to
finding what is the minimum depth of the tree (or multiple trees)
across which the stream is distributed. This problem has been
thoroughly addressed in \cite{sigmetrics08}, under the general
assumption that a stream may be subdivided into sub-streams
(delivered across different paths), and that each node may upload
information to a given maximum number of children. For instance, if
we assume no restriction on the number of children a node may upload
to, then it is proven in \cite{sigmetrics08} that a tree depth equal
to two is always sufficient. This is indeed immediate to understand
and visualize in the special case of all links with a ``sufficient''
amount of available upload bandwidth - see figure \ref{fig:2a} for a
constructive example\footnote{
        In this example, the amount of available
        upload bandwidth is ``sufficient'' in the sense that the source node has a
        bandwidth at least equal to the stream bit rate $R$, while
        each peer node has a bandwidth at least equal to
        $(N-1)\cdot R /N $, being $N$ the number of peer nodes
        composing the overlay. As shown in \cite{sigmetrics08}
        the same result holds under significantly less restrictive
        assumptions on the available bandwidth.
        }.

At this stage, it should be clear that, in the context of flow-based
systems, as long as some feasibility conditions are met (see e.g.
\cite{ross07}), the bandwidth available on each link plays a limited
role with respect to the delay performance achievable. This is
clearly seen by looking again at figure \ref{fig:2a}: if for
instance we double the bandwidth available on each link, the delay
performances do not change (at least until the source is provided
with a large enough amount of bandwidth to serve all peers in a
single hop).

\subsection{Delay in chunk-based systems}
\label{ss:chunk}%
Chunk-based systems have a key difference with
respect to flow-based systems: the streaming information is
organized into chunks whose size is significantly greater than IP
packets. Since a peer must complete the reception of a chunk before
forwarding it to other nodes (i.e. chunks are delivered in a
store-and-forward fashion), the obvious consequence is that delay
performance are mostly affected by the chunk transmission time.
Thus, in terms of delay performance, the behavior of chunk-based
systems is opposite to the one of flow-based systems. Not only chunk
transmission times cannot be neglected anymore with respect to
link-level delays (propagation and underlying network queueing), but
also we can safely assume that in most scenarios any other delay
component at the link-level has negligible impact when compared with
the chunk transmission times. This consideration can be restated as:
the delay performances of chunk-based systems do not depend on the
sum of the delays experienced while traveling over an overlay link,
but depend on the sum of the delays experienced while {\em crossing
a node}.

From a superficial analysis, one might argue that the overall delay
optimization problem does not change. In fact, the transmission
delay of a chunk at a given node could be attributed to the overlay
link over which the chunk is being transmitted, and, also in this
case, the optimization could be stated as a minimum path cost
problem.

\begin{figure}[t]
    \centering
    \includegraphics[scale=0.65]{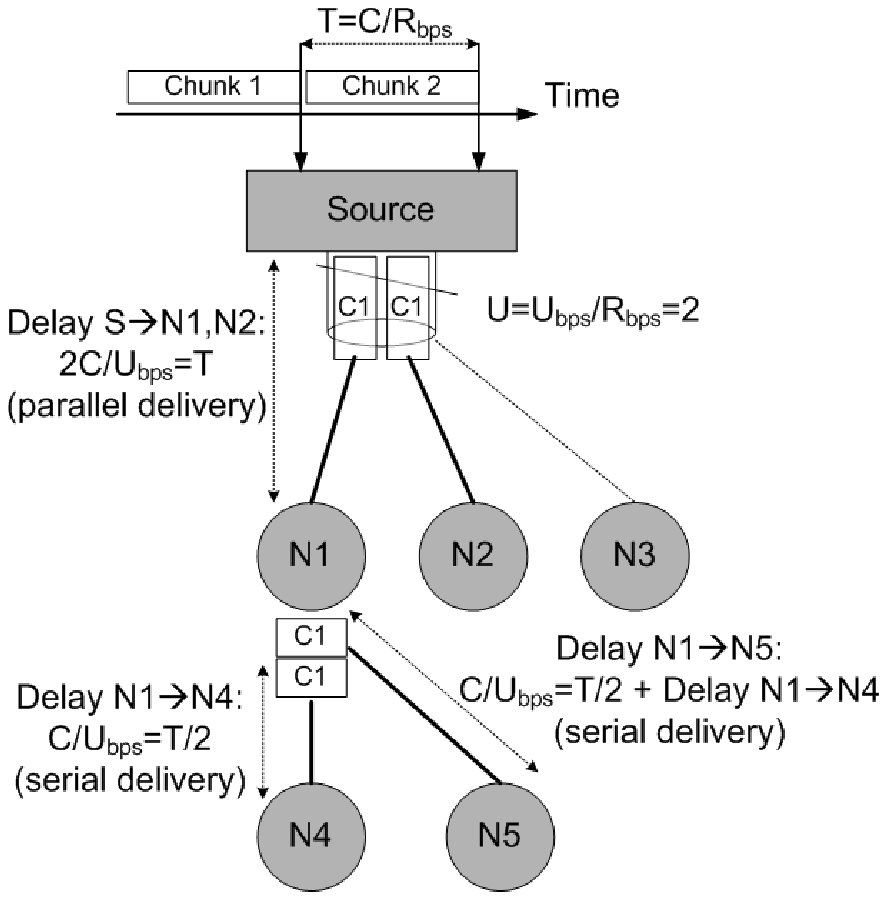}
    \caption{Delay components and constraining issues in chunk-based systems.}
    \label{fig:2b}
    \vspace{-0.8cm}
\end{figure}

However, a closer look reveals that this is not at all the case. The
reasons are manifold and can be illustrated with the help of figure
\ref{fig:2b}. In this figure, and consistently throughout the paper,
we rely on the following notation. $C$ is the chunk size (in bit);
$R_{bps}$ is the streaming constant bit rate (in bps). $T =
C/R_{bps}$ is the chunk ``inter-arrival'' time at the source, being
such arrival process a direct consequence of the segmentation into
chunks done at the source: a new chunk will be available for
delivery only when $C$ information bits, generated at rate
$R_{bps}$, are accumulated (see top of figure \ref{fig:2b}).
$U_{bps}$ is the available upload bandwidth, assumed to be the same
for all network nodes, including the source (homogeneous bandwidth
conditions). $U = U_{bps}/R_{bps}$ is the normalized upload
bandwidth of each node with respect to the streaming bit rate. In
this paper, for simplicity, we consider the case of $U$ integer
greater or equal than 1, i.e. $U_{bps}$ being either equal or a
multiple of $R_{bps}$. The {\em minimum} transmission time for a
chunk is equal to $T^* = C/U_{bps} = T/U$; this is true only if the
whole upload bandwidth $U_{bps}$ is used to transmit {\em a single
chunk to a single node}. Moreover, we rely on the common simplifying
assumption, in overlay P2P systems, that the only bandwidth
bottleneck is the uplink bandwidth of the access link that connects
the peer to the underlying network (the downlink bandwidth is
considered sufficiently large not to be a bottleneck - this is
common in practice, due to the large deployment of asymmetric access
links  - e.g., ADSL).

The first reason why the overall delay optimization problem can not
be stated as a minimum path cost problem in the case of chunk-based
systems is the sharing of the available upload bandwidth $U_{bps}$
across multiple overlay links. As a consequence, i) it is not
possible to {\em a priori} associate a constant delay cost to
overlay links originating from a given node, ii) the delay
experienced while transmitting a chunk depends on the fraction of
the bandwidth that the node is dedicating to such transmission. For
instance, figure \ref{fig:2b} shows that the source node is
transmitting a given chunk in parallel to two nodes; as such, the
transmission delay is $C/(U_{bps}/2)$. If the source were
transmitting the chunk only to node 1, this delay would be halved.

The second reason is that the transmission time may not be the {\em
only} component of the overall chunk delivery delay. This is
highlighted for the case of node N1. After receiving chunk 1, node
N1 adopts the strategy of {\em serializing} the delivery of chunk 1
to nodes N4 and N5. On the one side, in both cases the chunk will be
transmitted in the same time, namely $C/U_{bps}$; this is the
minimum transmission time for a chunk, as all the available
bandwidth is always dedicated to a single transmission
. On the other side, the time elapsing between the instant at which
the chunk is available at node N1 and the instant at which the chunk
is received by node N5 is greater than the transmission time, as it
includes also the time spent by node N1 while transmitting the chunk
to node N4.

The third and final aspect which characterizes chunk-based systems
in a {\em streaming} context is that there is a tight constraint
which relates the number of peer nodes that can be {\em
simultaneously} served and the available upload bandwidth. If we
look back flow-based systems in figure \ref{fig:2a}, we see that
only practical implementation issues may impede the source node to
arbitrarily subdivide the stream into sub-streams, and the tree
depth may be indeed trivially optimized by using as many sub-streams
as the number of nodes in the network. On the contrary, in
chunk-based systems, the number of nodes that can be served is no
more a ``free'' parameter, but it is tightly constrained by the
stream rate and the available upload bandwidth. This fact can be
readily understood by looking at the source node in the example
illustrated in figure \ref{fig:2b}. Due to their granularity, new
chunks are available for delivery at the source node every
$T=C/R_{bps}$ seconds. Hence, in order to keep the distribution of
chunks balanced (i.e., to avoid introducing delays with respect to
the time instant at which chunks are available at source and to
privilege specific chunks by giving them extra distribution time),
the source node must complete the delivery of every chunk before the
next new chunk is available for the delivery (i.e. within $T$
seconds). This implies that the source node cannot deliver a single
chunk to more than $U$ nodes, being $U=U_{bps}/R_{bps}$ the ratio
between the upload bandwidth and the streaming rate\footnote{
        A similar conclusion can be drawn for other nodes as well. Moreover,
        we remark that this conclusion holds even when chunks are serially
        delivered, like in the case of node N1.
        }.

\section{Stream diffusion metric: a delay-related fundamental bound}
\label{s:bound}

Let $\mathcal{P}$ be the set of all peers which compose a P2P
streaming network, and let $P = \left|\mathcal{P}\right|$ be the
cardinality of such network. Let $p \in \mathcal{P}$ be a generic
peer in the network. Since the streamed information is organized
into subsequently generated chunks, $p$ is expected to receive all
these chunks with some delay after their generation at the source.
Let us define with $d(c,p)$ the specific interval of time elapsing
between the generation of chunk $c$ ($c=1,2,3, \cdots$) at the
source, and its completed reception at peer $p$. In most generality,
different chunks belonging to the stream may be delivered through
different paths. This implies that $d(c,p)$ may vary with the chunk
index $c$. Let
\[ D(p) = \max_{c} d(c,p) \]
be the maximum delay experienced by peer $p$ among all possible
chunks.

To characterize the delay performance of a whole P2P streaming
network, we are interested in finding the maximum of the delay
experienced across all peers composing the network, i.e.:
\[ D\left( \mathcal{P} \right) = \max_{p \in \mathcal{P}} D(p) \]
We refer to this network-wide performance metric as {\em absolute
network delay}. However, for reasons that will be clear later on,
this performance metric does not yield to a convenient analytical
framework. Thus, we introduce an alternative delay-related
performance metric, which we call {\em stream diffusion metric}.
This is formally defined as follows:
\[ N(t) = \left|\mathcal{P}_t\right| \ \ \ \ \
                    {\rm where} \ \ \ \ \
                    \mathcal{P}_t = \{p \in \mathcal{P} : D(p) \leq t \} \]
In plain words, $N(t)$ is the number of peers that may receive each
chunk in at most a time interval $t$ after its generation at the
source.

The most interesting aspect of the stream diffusion metric $N(t)$ is
that it can be conveniently applied also to networks composed of an
infinite number of nodes (for such networks, obviously, the absolute
network delay $D\left( \mathcal{P} \right)$ would be infinite).
Moreover, for finite-size networks, it is straightforward to derive
the absolute network delay from the stream diffusion metric. Since
$N(t)$ is a non-decreasing monotone function of the continuous time
variable $t$ and it describes the number of peers that may receive
the whole stream within a maximum delay $t$, for a finite size
network composed of $P$ peers the value of $t$ at which $N(t)$
reaches $P$ is also the maximum delay experienced across all peers.
The formal relation between the absolute network delay and the
stream diffusion metric is hence
\[ D\left( \mathcal{P} \right) = \min \{t : N(t) = P \} \]

\subsection{The bound on $N(t)$}

Before stating the bound, we need to provide some preliminary
notation.

Let $F_k(i)$ be the $k$-step Fibonacci sequence defined as follows:
\begin{equation}
F_k(i) = \left\{
        \begin{array}{c l}
                0 & \mbox{if } i \leq 0 \\
                1 & \mbox{if } i=1  \\
                \sum_{j=1}^{k} F_k(i-j) & \mbox{if } i > 1   \\
        \end{array}
        \right.%
\label{e:fib}
\end{equation}
Let $S_k(n)$ be a new sequence defined as the sum of the first $n$
non-null terms of the $k$-step Fibonacci sequence, i.e.,
\begin{equation}
        S_k(n) = \left\{
                \begin{array}{c l}
                        0 & \mbox{if } n \leq 0 \\
                        \sum_{i=1}^n F_k(i) & \mbox{if } n > 0
                \end{array}
                \right.%
\label{e:fib-summa-def}
\end{equation}

Let us assume that propagation delays and queueing delays
experienced in the underlying physical network because of congestion
are negligible with respect to the minimum chunk transmission time
$T^*=C/U_{bps}$, namely the time needed to transmit a chunk by
dedicating, to such transmission, {\em all} the upload capacity of a
node. In what follows, we measure the time using, as time unit, the
value $T^*$ above defined.

We can now state the following theorem on the upper bound of $N(t)$.
\newtheorem{theorem}{{\bf Theorem}}
\setcounter{theorem}{0}
\begin{theorem}
\label{th:1}%
In a P2P chunk-based streaming system where all peer nodes have the
same normalized upload capacity $U=U_{bps}/R_{bps}$ (assumed integer
greater or equal than 1) and $k$ overlay neighbors to delivery
chunks to, the stream diffusion metric is upper bounded by
\begin{equation}
\overline{N}(t) = \sum_{j=1}^{U} S_k(t-j+1) \label{e:n-upper-bound}
\end{equation}
for integer values of $t$ (i.e. multiple of $T^*$) while, for non
integer values of $t$, $\overline{N}(t) = \overline{N}(\lfloor t
\rfloor)$ must be considered.
\end{theorem}

The proof of Theorem \ref{th:1} is omitted for reasons of space. We
refer the reader to \cite{tech-rep} for the full details. We only
observe that the proof is based on the following property: the
minimum amount of time elapsing between the time instant at which a
peer receives a chunk and the time instant at which it has
transmitted the received chunk to {\em i}, $i \in \{1,2,\dots,k\}$,
of its $k$ neighbors is lower bounded by $i$, and this is achieved
if and only if the chunk transmission is serialized. In other words,
the bound in (\ref{e:n-upper-bound}) may be achieved only by
serializing chunk transmissions.

\begin{figure*}[htbp]
    \centering
    \includegraphics[scale=0.75]{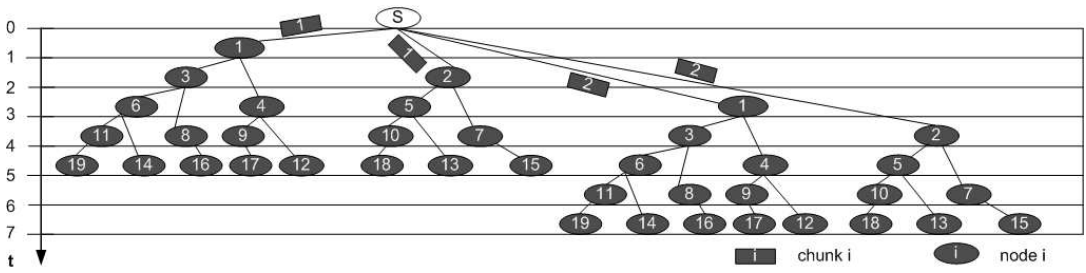}
    \caption{Overlay tree resulting in the case $k=U=2$. 
                }
    \label{f:serial-tree}
    \vspace{-0.6cm}
\end{figure*}
\subsection{Asymptotic closed form expressions for the bound on $N(t)$}
Thanks to the asymptotic expression of $k$-step Fibonacci Sums,
which has been derived in \cite{tech-rep}, equation
(\ref{e:n-upper-bound}) can be more conveniently expressed in the
following asymptotic closed form:
\begin{equation}
\begin{array}{c}
\overline{N}(t) = \displaystyle \sum_{j=1}^{U} S_k(t-j+1) \approx
\sum_{j=1}^{U} \frac{\phi_k \cdot
\phi_k^{t-j+1}}{(\phi_k-1)Q_k(\phi_k)}+ \\
\displaystyle - \sum_{j=1}^{U} \frac{1}{k-1} = \frac{\phi_k^2
(1-\phi_k^{-U})}{Q_k(\phi_k)(\phi_k-1)^2} \cdot \phi_k^t -
\frac{U}{k-1}
\end{array}
\label{e:n-approx-upper-bound}
\end{equation}
where i) $\phi_k$ represents the so said $k$-step Fibonacci constant
and it is the only real root with modulo greater than $1$ of the
characteristic polynomial $P_k(x) = x^k - x^{k-1} - x^{k-2} - \cdots
- x - 1$ of the $k$-step Fibonacci sequence, and ii) $Q_k(x)$ is a
suitable polynomial about which more details can be found in
\cite{tech-rep}.

For the convenience of the reader, the first few values of the
Fibonacci constants are
$\phi_2=1.61803,\phi_3=1.83929,\phi_4=1.92756,\phi_5=1.96595,\phi_6=1.98358$,
while the first few values of the terms $Q_k(\phi_k)$ are
$Q_2(\phi_2)=2.23607,Q_3(\phi_3)=2.97417,Q_4(\phi_4)=3.40352,Q_5(\phi_5)=3.65468,Q_6(\phi_6)=3.80162$.

The derived bound explicitly accounts for the fact that each node at
most can feed $k$ neighbors. If this restriction is removed, we
obtain a more simple and immediate expression (see \cite{tech-rep}
for more details)
\begin{equation}
\overline{N}(t)  = \sum_{j=1}^{U} S_\infty(t-j+1) =
            \sum_{j=1}^{U} 2^{t-j} = 2^t (1-2^{-U})
\label{e:n-bound-infty}
\end{equation}

\section{Attaining the bound}
\label{s:algo}

The provided bound offers only limited insights on how chunks should
be forwarded across the overlay topology. Specifically, the bound
clearly suggests that delay performances are optimized only if
chunks are serially delivered towards the neighbor nodes, but does
not make any assumption on which specific paths the chunks should
follow, or in other words, which overlay topologies should be used.
We now show that, to attain the performance bound, peer nodes have
to be organized according to i) an overlay unbalanced tree if $k=U$,
ii) multiple overlay unbalanced trees if $k>U$ and multiple of $U$
(generalization to arbitrary integer values of $k$ being
straightforward).

\begin{figure*}[htbp]
    \centering
    \includegraphics[scale=0.75]{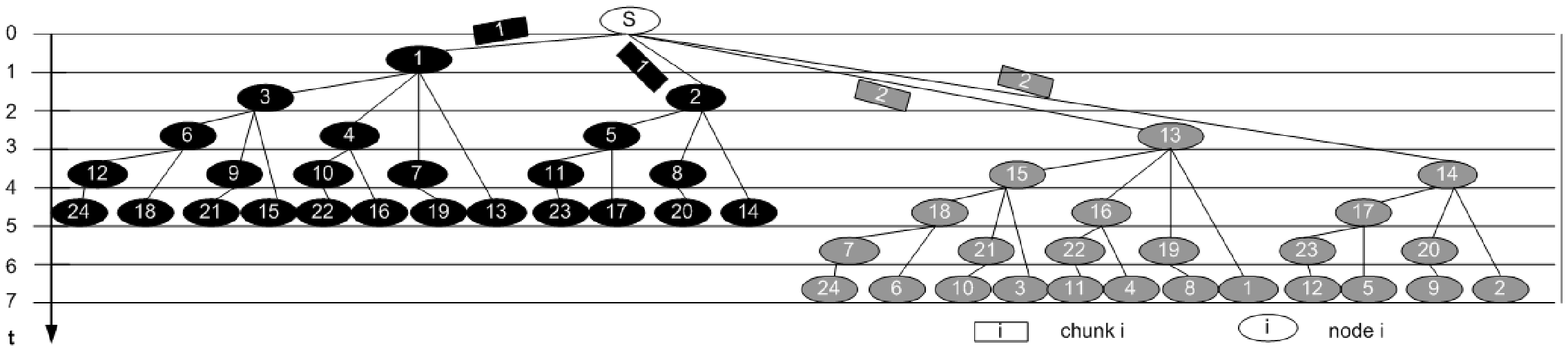}
    \caption{Chunk distribution over multiple trees for the case $U=2,k=4$.}
    \label{f:serial-forest}
    \vspace{-0.6cm}
\end{figure*}
\subsection{Case $k = U$: unbalanced tree}
\label{ss:tree}%
When the number of neighbor nodes $k$ is equal to the normalized
upload capacity $U$, the source node can deliver each chunk to {\em
all} its $k$ neighbors before a new chunk arrives. As such, the
source node can repeatedly apply a round-robin scheduling policy
during the time interval $T=U T^*$, which elapses between the
arrivals of consecutive chunks. Specifically, in the first $T^*$
seconds it can send a given chunk to a given node, say peer $N_1$,
then send the chunk to peer $N_2$, and so on until peer $N_k$. If
this policy is repeated for every chunk, the result is that any
neighbor of the source also receives a new chunk every $T=U T^*$
seconds. Hence, each neighbor of the source may apply the same
scheduling policy with respect to its neighbors, and so on. As a
consequence, every node in the network receives chunks from the same
parent, and in the original order of generation: in other words,
chunks are delivered over a tree topology.

The operation of the above described chunk distribution mechanism is
depicted in figure \ref{f:serial-tree}, which refers to the case
$U=k=2$ and a network composed of $19$ nodes. In this figure the
source is denoted with an ``S''. The nodes and the chunks are
progressively indexed starting from $1$. Going from the upper part
of the figure to its lower part, we see how the first two chunks are
progressively distributed starting from the source; the time since
the start of the transmission, measured in time units, until time
instant $t=7$ is reported on the left side of the figure. The tree
on the left hand side of the figure distributes the first chunk,
while the tree on the right hand side of the figure distributes the
second chunk. In more detail, since the first chunk is assumed to be
available for transmission at the source at time instant $t=0$, the
source starts transmitting the first chunk to node 1 at $t=0$ and
after finishing this transmission, i.e at $t=1$, it sends the first
chunk to node 2, in series. In its turn, node 1 sends the first
chunk first to node 3 and then to node 4, in series, and so on.
Likewise, node 2 sends the first chunk first to node 5 and then to
node 7, in series, and so on. As regards the second chunk, the
source starts transmitting it to node 1 at time $t=2$, exactly when
that chunk is available for the transmission. After finishing
transmitting the first chunk to node 1, the source sends the same
chunk to node 2, in series. In their turn, node 1 and 2 distribute
the second chunk in same manner as the first chunk, i.e. sending the
second chunk in series first to nodes 3 and 5 respectively, and then
to nodes 4 and 7 respectively.

It is to be noted that, even if two distribution trees are depicted
in figure \ref{f:serial-tree}, actually there is only one
distribution, which repeats itself for each chunk with period
$k=U=2$. In other words, a given node receives all chunks through
the same path. It is also interesting to note that the tree formed
in figure \ref{f:serial-tree} is unbalanced in terms of number of
hops. For instance, the first chunk reaches node 19 at time $t=5$
after crossing nodes 1,3,6 and 11. Conversely, the same chunk
reaches node 15, again at time $t=5$, after crossing nodes 2 and 7.
The unbalancing in terms of number of hops is a consequence of the
fact that the proposed approach achieves equal-delay
source-to-leaves paths, and that the time in which a chunk waits for
its transmission turn at a node (because of serialization)
contributes to such path delay.

We are now in condition to evaluate the stream diffusion metric
$N(t)$. To this end, let us introduce $n(i)$ as number of new nodes
that complete the download of a chunk exactly $i$ time units after
the generation of that chunk at the source node, in such a way that
$N(t)$ can be assessed according to the equation $N(t)=\sum_{i=1}^t
n(i)$. With reference to figure \ref{f:serial-tree}, $n(1)=1$ (node
1), $n(2)=2$ (nodes 2 and 3), $n(3)=3$ (nodes 4, 5 and 6), $n(4)=5$
(nodes 7, 8, 9, 10 and 11), $n(5)=8$ (nodes 12, 13, 14, 15, 16 and
17). Thus, $N(t) = 19$, which is equal to the performance bound
$\overline{N}(t)$ evaluated at $t=5$. To generalize the evaluation
of $n(i)$, we observe that only the nodes which have completed the
download of a chunk exactly after $i-1,i-2,i-3,\cdots,i-k$ since the
generation of that chunk have still children to be served, whereas
nodes that have completed the download of that chunk with a delay
less than $i-k$ have already served all their $k$ children. As a
consequence, if we set $n(0)=1$ to take the children served by the
source into account, it results $n(i) = n(i-1) + n(i-2) + \cdots +
n(1) + n(0)$ for $i \leq U$ and $n(i) = n(i-1) + n(i-2) + \cdots +
n(i-k+1) + n(i-k)$ for $i > U$. It is then easy to evaluate the
sequence $n(i)$ for a given $k=U$ and to verify that $n(i)=F_k(i+1)$
and consequently $N(t)=\sum_{i=1}^t F_k(i+1)$. Easy algebraic
manipulations allow to turn the last equality into
$N(t)=\sum_{j=1}^k S_k(t-j+1)$, which guarantees the matching
between the stream diffusion metric of the described chunk
distribution mechanism and the performance bound $\overline{N}(t)$
for each value of $t$.

\subsection{Case $k > U$ and multiple of $U$: unbalanced multiple trees}
\label{ss:forest}%
When $k>U$, the source cannot deliver a chunk to all its $k$
neighbors, but only to a subset of $U$ peers. Hence, in principle,
it might distribute chunks through the same tree as discussed
before, and hence every peer in the network would use only $U$
neighbors out of the available $k$. However, the provided bound
assures that performance in the case $k>U$ are better than in the
case $k=U$. For instance, if $U=2$, the case $k=4$ outperforms the
case $k=2$ as follows:
\begin{center}
\begin{tabular}{|l|l l l l l l l l |} \hline
$t$                                         & 1 & 2 & 3 & 4  & 5  &
6  & 7  & \ldots \\ \hline $\overline{N}(t), k=2$  & 1 & 3 & 6 & 11
& 19 & 32 & 53 & \ldots \\ \hline $\overline{N}(t), k=4$  & 1 & 3 &
6 & 12 & 24 & 47 & 91 & \ldots \\ \hline
\end{tabular}
\end{center}

A thorough general explanation of how to design a mechanism which
attains the bound in the case $k>U$ and multiple of $U$ is complex
(for reasons that will emerge later on). Hence, in this paper we
limit ourselves to show how the bound may be achieved through the
simple example depicted in figure \ref{f:serial-forest}, which
refers to the case $U=2$ and $k=4$ and a network composed of $24$
nodes. The notation in this figure is the same as in figure
\ref{f:serial-tree}. As in the case $k=U=2$, at time $t=0$ the
source node receives chunk \#1 and serially delivers it to nodes 1
and 2. However, with respect to the case $k=U=2$, at time $t=2$,
when the source node receives chunk \#2, instead of sending it again
to nodes 1 and 2, it delivers that chunk to the remaining two
neighbors (nodes 13 and 14). This process is repeated for the
subsequent chunks, and specifically the odd-numbered chunks are
serially delivered to nodes 1 and 2, while the even-numbered ones
are serially delivered to nodes 13 and 14. As a consequence of this
operation of the source, each neighbor of the source i) receives
directly from the source only half chunks, ii) receives a new chunk
from the source every 4 time units. As such, neighbors of the source
have the necessary extra time to deliver each chunk they receive
from the source to all their $k=4$ neighbors. The same holds for the
remaining peer nodes. For instance, with regard to chunk \#1, node 1
serves that chunk to all its four neighbors (nodes 3, 4, 7 and 13)
in series. Node 2 serves instead chunk \#1 only to three neighbors
(nodes 5, 8 and 14) out of four available, since all nodes in the
network have already received chunk \#1 at $t=5$ and there are no
nodes to be served. In their turn, all nodes that have been served
by nodes 1 and 2, transmit chunk \#1 to their neighbors (unless
their neighbors have already received that chunk) in series, and so
on, until all nodes in the network receive chunk \#1. This allows
delivering chunk \#1 to 24 nodes in 5 time units, instead of the
previous 19 nodes.

It is to be noted that chunks are now distributed by means of two
distinct unbalanced trees, the left one for odd-numbered chunks and
the right one for even-numbered chunks, which repeat themselves with
period $k=4$. In general, the number of distribution trees is $k/U$,
where we use the assumption that $k$ is integer multiple of $U$.

We are now in condition to evaluate the stream diffusion metric
$N(t)$. As in the case $k=U$, let us introduce $n(i)$ as number of
new nodes that complete the download of a chunk exactly $i$ time
units after the generation of that chunk at the source node, in such
a way that $N(t)$ can be assessed according to the equation
$N(t)=\sum_{i=1}^t n(i)$. With reference to figure
\ref{f:serial-tree} and to the left hand side tree, $n(1)=1$ (node
1), $n(2)=2$ (nodes 2 and 3), $n(3)=3$ (nodes 4, 5 and 6), $n(4)=6$
(nodes 7, 8, 9, 10, 11 and 12), $n(5)=12$ (nodes 13, 14, 15, 16, 17,
18, 19, 20, 21, 22, 23 and 24). The amounts $n(i)$ take on the same
values even in the right hand side tree. Thus, $N(t) = 24$, which is
equal to the performance bound $\overline{N}(t)$ evaluated at $t=5$.
To generalize the evaluation of $n(i)$, we observe that, if $i \leq
U$, the source is still serving a given chunk; otherwise, the source
is already serving the next chunk. In addition, only the nodes which
have completed the download of a chunk exactly after
$i-1,i-2,i-3,\cdots,i-k$ since the generation of that chunk have
still children to be served, whereas nodes that have completed the
download of that chunk with a delay less than $i-k$ have already
served all their $k$ children. As a consequence, if we set $n(0)=1$
to take the children served by the source into account, it results
$n(i) = n(i-1) + n(i-2) + \cdots + n(1) + n(0)$ for $i \leq U$,
$n(i) = n(i-1) + n(i-2) + \cdots + n(2) + n(1)$ for $U < i \leq k$
and $n(i) = n(i-1) + n(i-2) + \cdots + n(i-k+1) + n(i-k)$ for $i >
U$. It is then easy to evaluate the sequence $n(i)$ for a given pair
of $k$ and $U$ values and to verify that $n(i)=F_k(i) + F_k(i-1) +
\cdots + F_k(i-U+1)$ and consequently $N(t) = \sum_{i=1}^t
\sum_{j=1}^U F_k(i-j+1)$. Easy algebraic manipulations allow to turn
the last equality into $N(t)=\sum_{j=1}^k S_k(t-j+1)$, which
guarantees the matching between the stream diffusion metric of the
described chunk distribution mechanism and the performance bound
$\overline{N}(t)$ for each value of $t$.

Before concluding the description of the case $k>U$ and multiple of
$U$, we finally observe that a peer node needs to be part of all the
$k/U$ trees in order to properly receive the full stream. This leads
to a complex issue which we call the ``tree intertwining problem'',
that is: how nodes should be placed in every tree so that the
different role of a node in every considered tree does not lead to
sharing the node's upload capacity among the different trees (and
hence to performance impairments with respect to the bound's
prediction, or even congestion). This can be more easily illustrated
through the following example. Let us first consider node 5. In the
left (odd-numbered) tree, node 5 is in charge of serving two
neighbors, namely 11 and 17. If node 5 were used by the right
(even-numbered) tree in place of node 15, it would also have to
forward even-numbered chunks to three additional neighbors, thus
breaking the assumption that a node has at most $k=4$ neighbors. The
problem is actually more complex, as we can understand by
considering the following second case. In the odd-numbered tree,
node 2 has to serve three nodes, namely nodes 5, 8, and 14. At a
first glance, we might conclude that node 2 can be also used by the
even-numbered tree provided that it is placed in a position of the
tree that requires the node to serve only a single node. However,
this is not the case. In fact, let us assume to replace node 7 in
the even-numbered tree with node 2. This implies that node 2 would
be required to deliver an even-numbered chunk to node 24 at every
time instant $t=6+4n$. However, node 2 is required by the left tree
to deliver an odd-numbered chunk at instants of time $t=2+4n,
t=3+4n$, and $t=4+4n$. Thus, since $6+4n=2+4(n+1)$, node 2 should
simultaneously deliver an odd-numbered chunk to node 5, and an
even-numbered chunk to node 24, which would not allow reaching the
bound.

Unfortunately, the ``intertwining problem'' for unbalanced trees can
not be solved by letting interior nodes of a given tree play the
role of leaves in the remaining trees\footnote{
        This is instead the solution when parallel transmission and, as
        consequence, balanced trees are used \cite{ptree}, being trivial to show that the number of
        leaves in a tree of fan-out $k$ is greater than $(k-1)$ times the number of non-leaf
        nodes.
        }.
However, we proved in \cite{tech-rep} that i) the tree-intertwining
problem can be solved via exhaustive search for arbitrary $U$ and
$k$ and for any network size for which the bound $\overline{N}(t)$
is attainable, and that ii) there exists a constructive approach
which allows finding one of the many possible solutions without
relying on exhaustive search. Since this proof is complex and it
requires significant extra space and technical elaboration, we refer
the interested readers to \cite{tech-rep} for the details.

\section{Performance Evaluation}
\label{s:perfo}%

\begin{figure}[ht]
    \centering
    \includegraphics[scale=0.65]{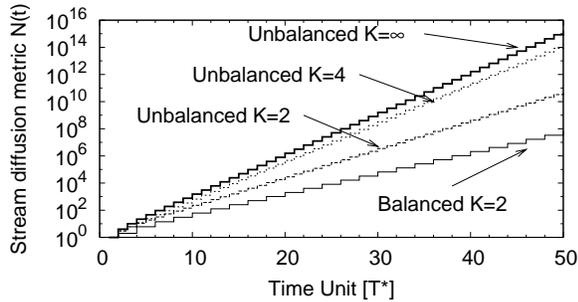}
    \caption{Stream diffusion metric $N(t)$ as a function of $T^*$ in a $U=2$ bandwidth scenario, for
    $k=2$ (balanced and unbalanced tree), $k=4$ (two unbalanced
    trees), and $k=\infty$ (infinite unbalanced trees).}
    \label{f:change_groups}
\end{figure}

Figure \ref{f:change_groups} plots the stream diffusion metric
$N(t)$ as a function of $T^*$ in a $U=2$ bandwidth scenario, for a
single unbalanced tree ($k=2$), two unbalanced trees ($k=4$),
infinite unbalanced tree ($k=\infty$) and a single {\em balanced}
tree ($k=2$ and parallel transmissions).

The first important observation about figure \ref{f:change_groups}
regards the impact of the number of neighbor nodes $k$ on the stream
diffusion metric bound. The figure shows that there is a significant
improvement when moving from the case $k=U=2$ of single tree to that
of multiple trees. Interestingly (but expected, as the Fibonacci
constants $\phi_k$ increase only marginally when $k$ becomes large),
the advantage in using more than a few trees is limited: this is
especially important if an algorithm is designed to mimic the
unbalanced multiple tree operation, as complexity (i.e. signalling
burden) increases with $k$.

The second important observation regards the improvement brought
about by serializing the transmissions (and hence unbalanced trees)
with respect to parallel chunk transmissions (and hence balanced
trees). The figure shows that the performance improvement is
significant: in the case $k=2$ the stream diffusion metric $N(t)$
for serial chunk transmissions (i.e., the bound) is one order of
magnitude greater than for parallel chunk transmissions at $t=20$,
and three orders of magnitude at $t=50$.

\section{Related work}
\label{s:related}%
The literature abounds of papers proposing practical and working
distribution algorithms for P2P streaming systems; however very few
theoretical works on their performance evaluation have been
published up to now. As a matter of fact, due to the lack of basic
theoretical results and bounds, common sense and intuitions and
heuristics have driven the design of P2P algorithms so far.

The few available theoretical works mostly focus on the flow-based
systems, as they have been defined in subsection \ref{ss:flow}. In
such case, a fluidic approach is typically used to evaluate
performance and the bandwidth available on each link plays a limited
role with respect to the delay performance, which ultimately depend
on the delay characterizing a path between the source node and a
generic end-peer. This is the case in \cite{sigmetrics08} and
\cite{ross07}. Moreover, there are also other studies that address
the issue of how to maximize throughput by using various techniques,
such as network coding \cite{baochunli05} or pull-based streaming
protocol \cite{zhang07}.

This work differs from the previously cited ones mainly because it
focuses on chunk-based systems, for which discrete-time approaches
are most suitable than fluidic approaches. Surprisingly enough,
according to the best of our knowledge and our literature survey,
there is only one work \cite{snowball} where chunk-based systems are
theoretically analyzed. In more detail, the author of
\cite{snowball} derives a minimum delay bound for P2P video
streaming systems, and proposes the so called {\em snow-ball}
streaming algorithm to achieve such bound. Like the theoretical
bound presented in this paper, the bound in \cite{snowball}, that is
expressed in terms of delay in place of stream diffusion metric, can
be achieved only in case of serial chunk transmissions and it is
equivalent to the one that we found as a particular case when
$k\rightarrow\infty$. However, the assumptions under which such
bound has been derived in \cite{snowball} are completely different.
In fact, with reference to a network composed of $N=2^l$ nodes
excluding the source node, the proposed {\em snow-ball} algorithm
for chunk dissemination requires that i) the source node serves each
one of the $N=2^l$ network nodes with different chunks, ii) nodes
other than the source serve $l$ different neighbors. In other words,
the resulting overlay topology is such that i) the source node is
connected to all the $N$ network nodes, ii) nodes other than the
source have $log_2 N$ overlay neighbors. Due to this, our approach
may be definitely regarded as significantly different from the one
in \cite{snowball}. Differently from \cite{snowball}, we indeed
consider the case of limited overlay connectivity among nodes and we
show that organizing nodes in a forest-based topology allows to
achieve performance very close to the ones of the snow-ball case.

\section{Conclusions}
\label{s:conclu}%
In this paper we derived a theoretical performance bound for
chunk-based P2P streaming systems. Such bound has been derived in
terms of the stream diffusion metric, a performance metric which is
directly related to the end-to-end minimum delay achievable in a P2P
streaming system. The presented bound for the stream diffusion
metric depends on i) the upload bandwidth available at each node,
assumed homogeneous for all nodes, and ii) the number of neighbors
to transmit chunks to. k-step Fibonacci sequences play a fundamental
role in such a bound. The importance of the presented theoretical
bound is twofold: on the one hand, it provides an analytical
reference for performance evaluation of chunk-based P2P streaming
systems; on the other hand, it suggests some basic principles, which
can be exploited to design real-world applications. In particular,
it suggests i) the serialization of chunk transmissions, and ii) the
organization of chunks in different groups so that chunks in
different groups are spread according to different paths. 

\end{document}